\documentclass[a4paper]{mem}
\usepackage{natbib}
\usepackage{graphicx}
\usepackage[a4paper]{hyperref}
\idline{73}{23}

\begin{document}

   \title{Low-Mass Companions to Solar-Type Stars}

   \author{Stanimir Metchev
          \and
          Lynne Hillenbrand}

   \offprints{Stanimir Metchev, \email{metchev@astro.caltech.edu}}

   \institute{Department of Astronomy,
California Institute of Technology, MC 105--24, Pasadena, California
91125, USA}

   \abstract{
We present preliminary results from a coronagraphic survey of young
nearby Sun-like stars
using the Palomar and Keck adaptive optics systems.  We have targeted
251 solar analogs (F5--K5) at 20--160 pc from the Sun,
spanning the 3--3000 Myr age range.  The youngest ($<$500~Myr)
$\approx$100 of these have been imaged with deeper exposures to search
for sub-stellar companions.
The deep survey is sensitive to brown-dwarf companions at separations
$>$0.5$\arcsec$ from their host stars, with sensitivity extending to
planetary-mass (5--15$M_{\rm Jup}$) objects at wider ($>$3$\arcsec$)
separations.  Based on the discovery of a number of new low-mass
($<$0.2$M_\odot$) stellar companions, we infer that their frequency 
at $>$20~AU separations (probed via direct imaging) may be greater 
($\approx$12\%) than that found from radial velocity surveys probing 
$<$4~AU separations \citep[$\approx$6\%;][]{mazeh_etal03}.
We also report the astrometric confirmation of the first sub-stellar
companion from the survey~--- an L4 brown dwarf at a projected 
distance of 44~AU from the $\approx$500~Myr-old star
HD~49197.  Based on this detection, we estimate that the frequency of 
sub-stellar companions to solar-type stars is at least 1\%, and possibly 
of order a few per cent.
   }
   \authorrunning{Metchev \& Hillenbrand}
   \titlerunning{Companions to Solar-Type Stars}
   \maketitle

\section{Introduction \label{sec_intro}}

High-contrast imaging searches for low-mass companions to nearby and/or
young stars have increased dramatically in number since the first
direct-imaging discovery of a brown dwarf around a main sequence star
\citep[Gl~229;][]{nakajima_etal95}.  Imaging with adaptive optics (AO)
 is a particularly powerful approach, as it provides
the high angular resolution ($\leq$0.1$\arcsec$) achievable at the
diffraction limit of large ground-based telescopes.
Young nearby stars are the most suitable targets for direct imaging of
sub-stellar companions.  At ages of 10--100 million years (Myr) the
expected brightness ratio in the near-IR between an object near the
deuterium-burning limit ($\approx$0.013$M_\odot \approx 13$ Jupiter masses
[$M_{\rm Jup}$]) and a solar-type star is $10^{-3}$--$10^{-4}$
\citep{burrows_etal97, baraffe_etal03}: within the dynamic range
attainable by modern AO systems at 0.5$\arcsec$--1$\arcsec$
from bright stars.  Therefore, AO surveys of young nearby stars allow
the direct imaging of brown-dwarf and even potential planetary-mass 
($<$13$M_{\rm Jup}$) companions at orbital separations comparable to 
the semi-major axis (40~AU) of Neptune's orbit in the Solar System.

The emergent picture from the first large-scale programs targeting nearby
($<$25~pc)
stars \citep[e.g.,][]{oppenheimer_etal01, mccarthy_zuckerman04} is that
brown dwarfs are rare at separations 10--1000~AU from main sequence stars
(frequency $\sim$1\%).  This has become known
as the ``brown dwarf desert at wide separations,'' following an analogy
with the radial-velocity (RV) brown dwarf desert at $<$3~AU
\citep{marcy_butler00}.
However, recent results from more sensitive space- and ground-based
imaging surveys, though not inconsistent with the initial estimates, all point
to a
somewhat higher (several per cent) frequency of sub-stellar companions
over 20--200~AU separations
\citep{potter_etal02, lowrance_etal03, neuhauser_guenther04}.  At even
wider orbital semi-major axes ($>$1000~AU), \citet{gizis_etal01} observe
that the sub-stellar companion frequency ($\sim$18\%) is comparable to
that of stellar companions.  It is
therefore possible that the brown dwarf desert at wide separations may at least
partially be an artifact of the limited performance of ground-based surveys.
To explore this possibility, we have conducted a large, sensitive survey
for low-mass companions to young stars, using the Palomar and Keck AO
systems.

\section{Survey Sample and Observations}

  \begin{figure*}
  \centering
  \resizebox{\hsize}{!}{\includegraphics[clip=true]{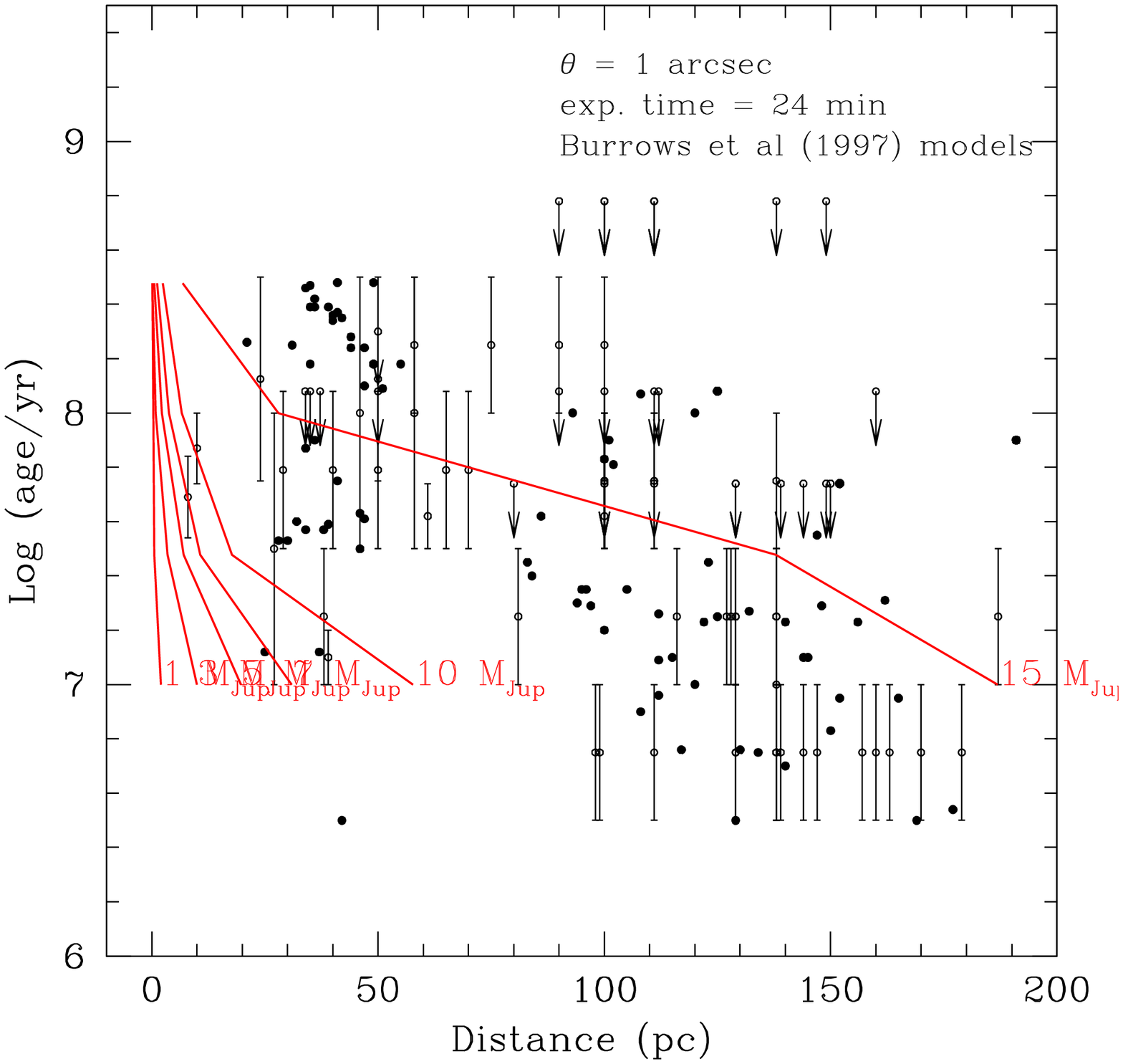}
  \includegraphics[clip=true]{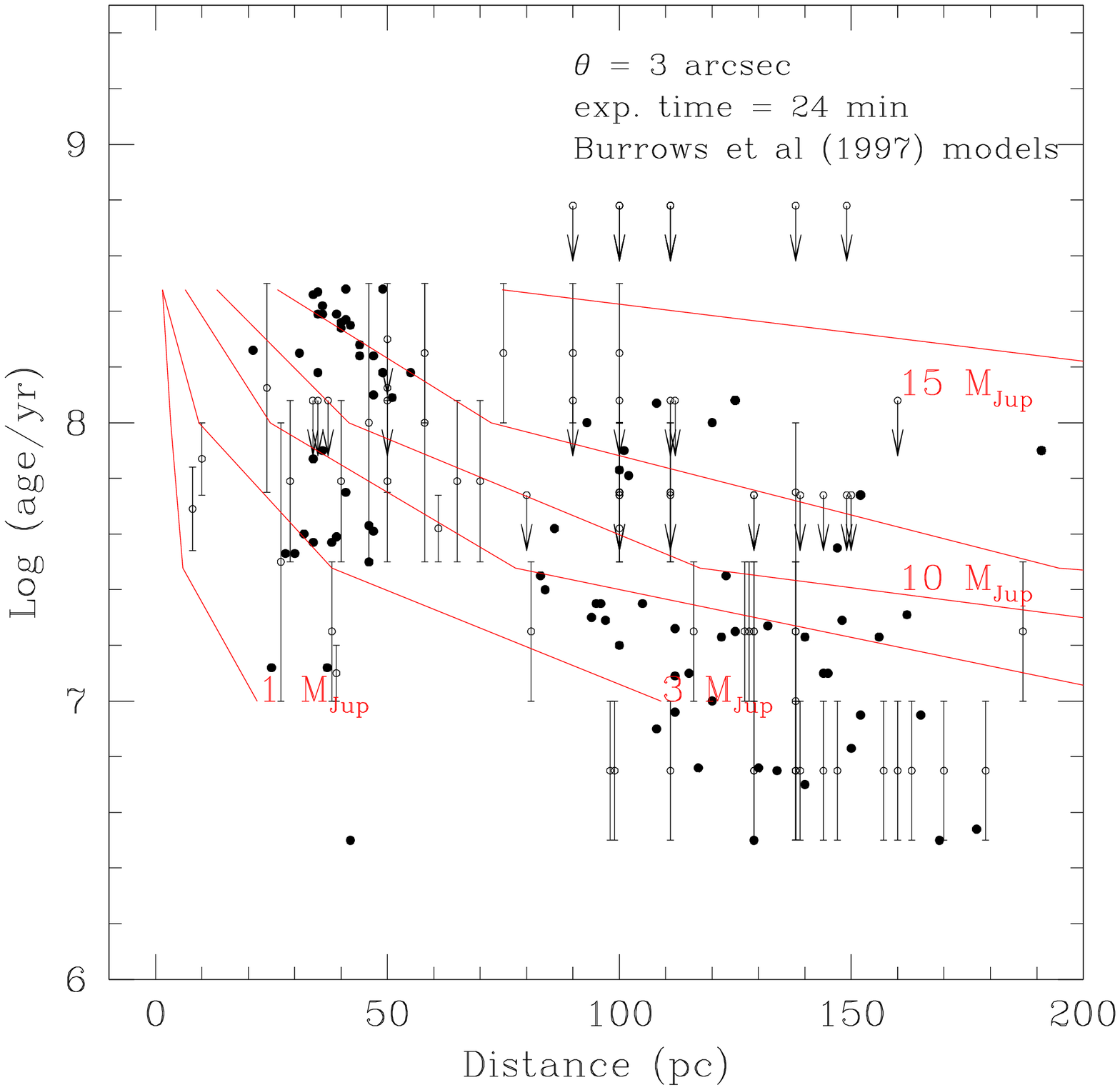}}
    \caption{The sample of $\approx$100 solar analogs in the deep survey 
	on an age vs.\ distance diagram.  Over-plotted (solid lines) are 
	the limiting survey sensitivities to brown dwarf companions of 
	a range of masses at angular separations of 1$\arcsec$ ({\sl left} 
	panel) and 3$\arcsec$ ({\sl right} panel) from their respective
	primaries.  That is, if a star residing below the $X 
	M_{\rm Jup}$ line in either of the panels had a $\geq X 
	M_{\rm Jup}$ companion at the corresponding angular separation
	(1$\arcsec$ or 3$\arcsec$), the companion should have been 
	detectable.  }  
       \label{fig_detlimits}
   \end{figure*}

The core of the survey targets $\approx$100 young ($<$500~Myr; median 
age 100~Myr) solar-type (F5--K5) stars
within 200~pc of the Sun (median distance 100~pc) using the high-order
AO system and coronagraph at the Palomar 5-m telescope \citep{troy_etal00}.  
A shallower non-coronagraphic survey, aimed at
determining the low-mass stellar multiplicity of older solar analogs,
encompasses an additional $\approx$150 stars.  The total number of solar 
analogs surveyed with the Palomar AO system is 251, spanning the 
age range 3--3000~Myr.  Follow-up astrometric and
spectroscopic observations are obtained at Palomar and/or Keck.

Although faint primary stars, 
such as M dwarfs or white dwarfs, offer more favorable
contrast for imaging sub-stellar companions, the sample was
limited to solar analogs only, since young F--G stars have remained
relatively unexplored by previous surveys because of their comparatively 
small numbers in the solar neighborhood.  
The source list is largely a sub-sample of the Sun-like stars targeted
by the Formation
and Evolution of Planetary Systems \citep[FEPS;][]{meyer_etal04} {\sl 
Spitzer} Legacy program.  We note that the FEPS sample is biased against 
equal-brightness binaries ($\Delta K_S<3$~mag and separation
$<$15$\arcsec$ in 2MASS), because they would complicate the analysis 
of {\sl Spitzer} data (PSF FWHM of 7$\arcsec$ at 70$\mu$m).
Nevertheless, a number of close binaries were found in the course of our
survey (Section~\ref{sec_multiplicity}) because of the higher
spatial resolution (0.10$\arcsec$ at $K_S$ band) allowed by AO compared
to seeing-limited observations.

A detailed description of the observing procedure is provided in 
\citet{metchev_hillenbrand04}.  Here we only state that imaging is
conducted at $K_S$ band (2.15$\mu$m) with total exposure times of 24~min per
star in the (deep) coronagraphic survey, and 10--50~s per star in the
(shallow) non-coronagraphic survey.  
Figure~\ref{fig_detlimits} shows the distribution of
the deep survey sample on an age vs.\ heliocentric distance diagram, with 
empirically-derived companion mass detection limits \citep[using sub-stellar
cooling models from][]{burrows_etal97} for 1$\arcsec$ ({\sl left} panel)
and 3$\arcsec$ ({\sl right} panel) separations.  Given the
dynamic range achievable with the Palomar AO system at $K_S$ 
($\approx$8.5~mag at
1$\arcsec$; $\approx$13.5~mag at 3$\arcsec$), we are sensitive to brown 
dwarfs above the deuterium-burning 
limit ($\sim$13 $M_{\rm Jup}$) around approximately half 
of the sample stars at 1$\arcsec$ separations, and around nearly all of 
the stars at 3$\arcsec$ separations.  Massive ($\geq$3$M_{\rm Jup}$) 
outer planets are also 
detectable around the youngest and the nearest of the solar analogs.

\section{Multiplicity \label{sec_multiplicity}}

   \begin{figure}
   \centering
   \resizebox{\hsize}{!}{\includegraphics{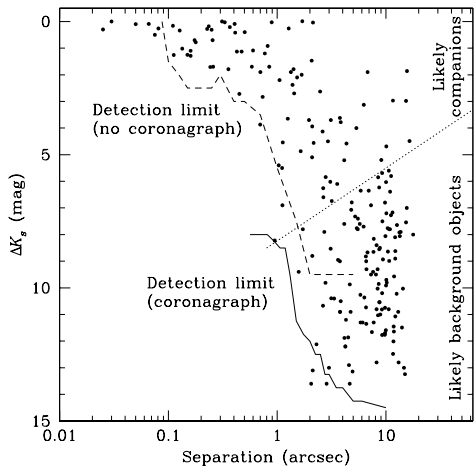}}
      \caption{Candidate companions from the first-epoch survey.
      $K_S$-band detection limits of the deep coronagraphic 
      (solid line) and shallow non-coronagraphic (dashed line) parts 
      of the survey are also shown.  Companions residing to the left 
      of the lines (beyond the $K_S$ sensitivity limits of the Palomar
      AO survey) were discovered either at
      $J$ band (where the PSF is only 0.05$\arcsec$ [FWHM], vs.\
      0.10$\arcsec$ at $K_S$) or during follow-up observations at
      Keck (where the higher dynamic range at $>$0.8$\arcsec$ is
      1--1.5~mag greater).
      The dotted line marks the empirically determined \citep{brandner_etal00}
      contamination rate from background objects at a Galactic latitude
      of $b\sim15\degr$.  At this latitude, objects residing below the 
      line are more likely to be unrelated background stars.  }
         \label{fig_proj}
   \end{figure}
 
First-epoch imaging ended in 2003, and the second-epoch
astrometric and spectroscopic follow-up is nearly ($\approx$90\%)
complete.  The first-epoch observations uncovered 207 candidate companions
within 15$\arcsec$ from 116 of the 251 stars in our overall sample 
(Figure~\ref{fig_proj}).  At the median age (100~Myr) and median 
$K$-band absolute magnitude ($M_K=3.5$~mag) of the stars in the deep 
sample, any candidate 
companion fainter than $\Delta K_S=6$~mag with respect to its primary is 
a possible brown dwarf, and any candidate fainter than $\Delta K_S=12$~mag 
is potentially below the deuterium-burning limit.
However, contamination from background sources is high, with the faintest
and most widely separated companions being most probably unrelated
background stars.  Figure~\ref{fig_proj} shows the loci of likely
companions and likely background stars, where they are separated 
by the dotted line delineating the empirically determined
\citep{brandner_etal00} background contamination rate for a Galactic
latitude of $b\sim15\degr$ (toward the Scorpius-Centaurus OB
association).  Given that the
majority of the stars in our sample reside at higher Galactic latitudes,
the line gives a conservative estimate (by 1--2 mag at moderate
latitudes) of the background contamination.

The astrometric analysis of the follow-up observations is currently
$\sim$20\% done, and an estimate of the overall survey completeness
has not been performed yet.  Therefore, at this stage we can only
present preliminary multiplicity results for the sample.
Nevertheless, for {\sl stellar} companions, these are not likely to differ
significantly from the final results, since the observed
fraction of stellar companions is less affected by background
contamination and incompleteness compared to that of sub-stellar
companions.  There are
78 candidate stellar binary systems in the locus of ``likely
companions'' in Figure~\ref{fig_proj}, where we have excluded the
companions discovered at separations smaller than the $K_S$-band 
diffraction limit (0.10$\arcsec$) of the Palomar AO system.  One out of 
78 is (spectroscopically) confirmed
to be sub-stellar (Section~\ref{sec_hd49197}), while the majority are
expected to be stellar.  For 47 of these systems the $K_S$-band flux 
ratio is $\Delta K_S<3.0$, and for 31 it is $\Delta K_S\geq3.0$.
Whereas the statistics of the $\Delta K_S<3.0$ bin are 
affected by the inherent sample bias against such binaries, the statistics
of the higher flux ratio bin are not.  At the median spectral type of the
sample (G5~V), a flux ratio of $\Delta K_S\geq3.0$ corresponds
approximately to a mass ratio of $q=M_2/M_1\leq0.2$ on the main
sequence.
Thus, we find that 12.4$\pm$2.2\% (assuming Poisson errors) of wide 
($\geq$20~AU) binaries have mass ratios $q\leq0.2$.  On the other hand, 
based on one confirmed brown dwarf companion among the 
100 stars in the deep sample, we find that the sub-stellar companion
frequency at wide orbital separations is likely of order 1\%.  However, neither
of these estimates is corrected for completeness, which, in the case of
the sub-stellar companion frequency, may boost the estimate by a
factor of several.

We put these numbers in the context of sensitive multiplicity surveys
targeting smaller orbital separations.  From a near-IR RV survey of
0.6--0.85$M_\odot$ primaries,
\citet{mazeh_etal03} find that $6.3\pm4.7\%$ of short-period
($<$3000-day) binaries have mass ratios $q\leq0.2$.  For the range of
primary masses in their sample,
the 3000-day period limit translates into a separation limit of
$<$4~AU.  On the other hand, \citet{marcy_butler00} estimate that $<$0.5\% of
Sun-like primaries have brown-dwarf companions orbiting within 3~AU.
Both estimates are somewhat lower than the
low-mass stellar and sub-stellar companion fractions found for the
widely-separated systems in our
survey.  We interpret this as evidence that the fraction of
low-mass stellar companions at wide ($>$20~AU) separations may be higher 
than that at small ($<$4~AU) separations.  The
tendency is also in agreement with the observed trend for an increasing 
frequency of sub-stellar companions at larger orbital semi-major axes
(Section~\ref{sec_intro}).
However, a robust conclusion will be possible only after the completion of the
common proper motion and completeness analyses of our entire sample.

\section{HD 49197B - a Young Ultra-Cool Brown Dwarf \label{sec_hd49197}}

   \begin{figure*}
   \centering
   \resizebox{\hsize}{!}{\includegraphics[clip=true]{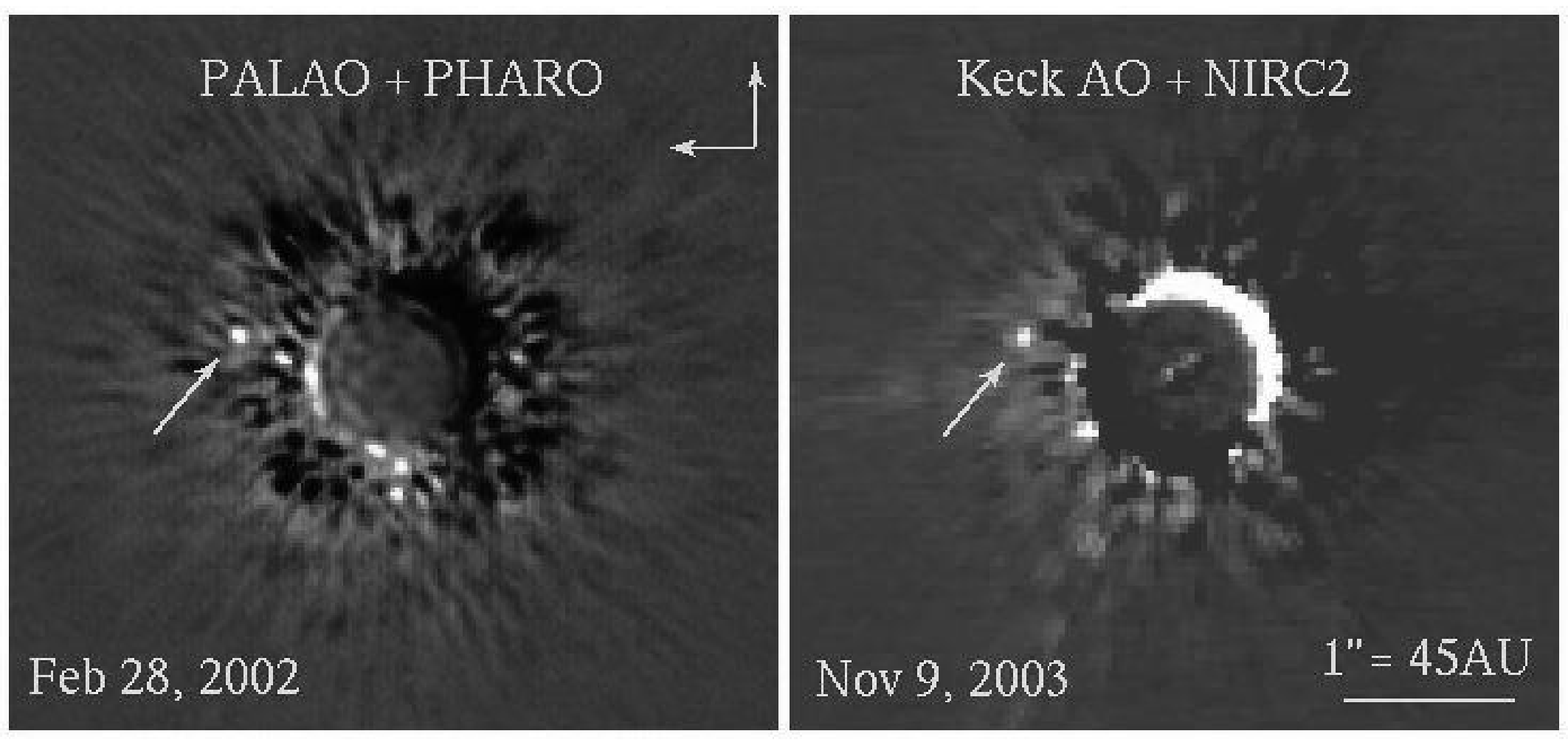}
   \includegraphics[clip=true]{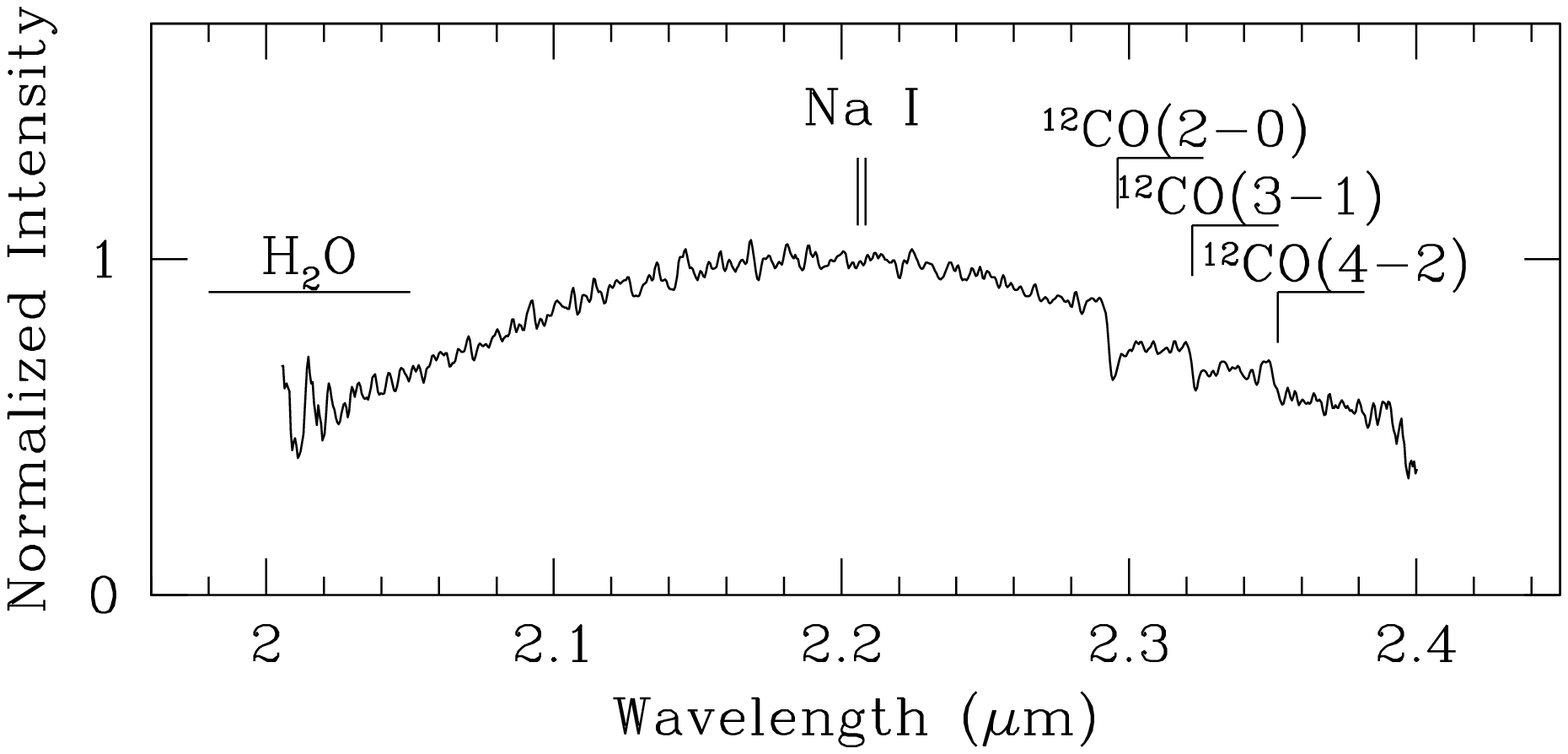}}
   \caption{A brown dwarf companion to the F5~V 260--790~Myr old
   star HD~49197 (45~pc, $K_S$=8.2~mag).  {\sl
   Left} panel: first- and second-epoch PSF-subtracted images confirming 
   its proper
   motion association with the primary.  HD~49197B is 8.2~mag fainter
   than the primary at $K_S$, and at a projected separation of 44~AU
   (0.95$\arcsec$).  {\sl Right} panel: $K$-band spectrum of HD~49197B, 
   showing H$_2$O and CO$_2$ absorption characteristic of cool dwarfs, and
   lack of Na~I absorption (not expected in mid-L dwarfs).  The spectral 
   type is determined to be 
   L4$\pm$1, and the inferred mass is 35--72$M_{\rm Jup}$.
     }
     \label{fig_hd49197b}
\end{figure*}

From the deep coronagraphic survey we have confirmed physical 
association between the components of one young stellar/sub-stellar binary: 
HD~49197A/B (Fig.~\ref{fig_hd49197b}).  Because the primary does
not belong to a known young moving group, its age can be constrained 
only approximately, based on chromospheric activity indicators and on
its photospheric lithium abundance.  From the strength of the Ca~H \& K core
emission measured in Keck/HIRES spectra, \citet{wright_etal04} assign an 
age of 525~Myr for HD~49197.  From our
own high-resolution optical spectra, we measure a lithium equivalent
width of 80~m\AA\ (Hillenbrand et al., in prep.), consistent with a
Pleiades-like \citep*[120~Myr;][]{stauffer_etal98} or older age.
Assuming that the Ca~H \& K age is accurate to 50\%, given the variation
in chromospheric activity of solar-type stars \citep{henry_etal96}, we
adopt an age range of 260--790~Myr for HD~49197A.  If the binary
components are co-eval, the mass of HD~49197B is 
$0.060_{-0.025}^{+0.012} M_\odot$, i.e., sub-stellar.

Given its association with a young main
sequence star, HD~49197B (L4$\pm$1~V) is a member of a very short list (5--6)
of confirmed {\sl young} ultra-cool (L or T) dwarfs.  Although the existence 
of ultra-cool dwarfs has also been reported toward young open clusters
\citep[e.g.,][]{zapatero_osorio_etal99, lucas_etal01,
lopez_marti_etal04}, their physical association with the corresponding
cluster, or their spectral classification based on near-IR
steam absorption indices, is still the subject of dispute
\citep[e.g.,][]{mcgovern_etal04, burgasser_etal04, luhman_etal03}.
Therefore, because of its rarity, the
physical properties of HD~49197 are of interest for constraining brown-dwarf
cooling models, and for projecting the spectroscopic
characteristics of even cooler objects of similar age~---
potential planets.

\section{Conclusion}

We have nearly completed a sensitive AO survey for low-mass companions
to young solar analogs.  Preliminary results indicate that 
low-mass stellar and sub-stellar ($M_2<0.2 M_\odot$) companions are 
more common at wide ($>$20~AU) than at small ($<$4~AU) orbital separations.  
The frequency of wide sub-stellar companions is found to be $\geq$1\%.
We also report the astrometric and spectroscopic confirmation
of the first brown dwarf companion to emerge from the survey:
HD~49197B.  Because of its youth, HD~49197B is a member of a short list
of known young ultra-cool dwarfs and is thus of interest as a benchmark
for the spectral classification and theoretical modeling of similar
objects in the future.  When completed, this survey will provide
factors of several better statistics on stellar multiplicity
across two dex of binary mass ratios, from 0.01--1.0, and over
10--1000~AU
separations, compared to previous surveys.  The result will be a much
more definitive estimate of the depth and extent of the brown dwarf desert.

\bibliographystyle{aa}

\end{document}